\documentclass[paper,notoc,12pt]{JHEP3}

\usepackage{amsmath}
\usepackage{graphicx}
\usepackage[sort&compress,numbers]{natbib}

\usepackage{psfrag}
\usepackage{epsf}

\allowdisplaybreaks[1]

\def\be{\begin{equation}}
\def\ee{\end{equation}}
\def\bea{\begin{eqnarray}}
\def\eea{\end{eqnarray}}

\def\Neqfour{{\cal N}=4}
\def\Neqone{{\cal N}=1}
\def\MHVb{$\overline{\text{MHV}}$}

\def\spa#1.#2{\left\langle#1\,#2\right\rangle}
\def\spb#1.#2{\left[#1\,#2\right]}
\def\lor#1.#2{\left(#1\,#2\right)}
\def\sand#1.#2.#3{%
\left\langle\smash{#1}{\vphantom1}^{-}\right|{#2}%
\left|\smash{#3}{\vphantom1}^{-}\right\rangle}
\def\sandp#1.#2.#3{%
\left\langle\smash{#1}{\vphantom1}^{-}\right|{#2}%
\left|\smash{#3}{\vphantom1}^{+}\right\rangle}
\def\sandpp#1.#2.#3{%
\left\langle\smash{#1}{\vphantom1}^{+}\right|{#2}%
\left|\smash{#3}{\vphantom1}^{+}\right\rangle}
\def\sandpm#1.#2.#3{%
\left\langle\smash{#1}{\vphantom1}^{+}\right|{#2}%
\left|\smash{#3}{\vphantom1}^{-}\right\rangle}
\def\sandmp#1.#2.#3{%
\left\langle\smash{#1}{\vphantom1}^{-}\right|{#2}%
\left|\smash{#3}{\vphantom1}^{+}\right\rangle}
\def\sandmm#1.#2.#3{%
\left\langle\smash{#1}{\vphantom1}^{-}\right|{#2}%
\left|\smash{#3}{\vphantom1}^{-}\right\rangle}
\def\spab#1.#2.#3{\sandmm#1.#2.#3}
\def\spbb#1.#2.#3.#4{\sandpm#1.{#2#3}.#4}
\newbox\charbox
\newbox\slabox
\def\s#1{{      
        \setbox\charbox=\hbox{$#1$}
        \setbox\slabox=\hbox{$/$}
        \dimen\charbox=\ht\slabox
        \advance\dimen\charbox by -\dp\slabox
        \advance\dimen\charbox by -\ht\charbox
        \advance\dimen\charbox by \dp\charbox
        \divide\dimen\charbox by 2
        \raise-\dimen\charbox\hbox to \wd\charbox{\hss/\hss}
        \llap{$#1$}
}}

\def\beqa{\begin{eqnarray}}
\def\eeqa{\end{eqnarray}}
\def\beq{\begin{equation}}
\def\eeq{\end{equation}}

\def    \br#1#2          {\mbox{$\langle #1 \, #2 \rangle$}}
\def    \sq#1#2          {\mbox{$\left[  #1 \, #2 \right]$}}
\def    \sap#1#2#3       {\mbox{$\langle #1 | #2 |#3  \rangle$}}
\def    \t#1#2#3         {\mbox{$s_{#1 #2 #3}$}}
\def    \s#1#2           {\mbox{$s_{#1 #2}$}}
\def    \sapp#1#2#3#4    {\mbox{$\langle #1 | (#2+#3) |#4  \rangle$}}

\def    \br(#1,#2)          {\mbox{$\langle #1 \, #2 \rangle$}}
\def    \sq(#1,#2)          {\mbox{$\left[  #1 \, #2 \right]$}}
\def    \t(#1,#2,#3)        {\mbox{$s_{#1 #2 #3} $}}
\def    \s(#1,#2)           {\mbox{$s_{#1 #2}$ }}
\def    \Pperp(#1)          {\mbox{$k_{#1 \perp}$ }}
\def    \PperpStar(#1)      {\mbox{$k_{#1 \perp}^*$ }}
\def    \Pperpt(#1)          {\mbox{$|k_{#1 \perp}|^2$ }}
\def    \x(#1)              {\mbox{$x_{#1}$ }}

\def\Del#1.#2.#3{\Delta(#1,#2;#3)}

\newcommand{\aab}[2]{\langle #1\, #2 \rangle}

\newcommand{\Split}{\mathrm{Split}}
\newcommand{\ssplit}{\mathrm{split}}

\author{
T.~G. Birthwright, E.~W.~N.~Glover, V.~V.~Khoze and P.~Marquard.\\
Department of Physics,
University of Durham,
Durham DH1 3LE,
U.K.\\
E-mail:  \email{T.G.Birthwright@durham.ac.uk, E.W.N.Glover@durham.ac.uk,
Valya.Khoze@durham.ac.uk, Peter.Marquard@durham.ac.uk}
}
\title{Collinear Limits in QCD from MHV Rules}

\abstract{
We consider multi-parton collinear limits of QCD
  amplitudes at tree level. Using the MHV formalism
  we specify the underlying analytic structure of the resulting
  multi-collinear splitting functions. We derive general results
  for these splitting functions
  that are valid for specific numbers of negative helicity
  partons and an arbitrary number of positive helicity partons (or vice versa).
 }

\keywords{QCD, Supersymmetry and duality, Hadronic colliders}

\preprint{DCPT/05/44, IPPP/05/22, hep-ph/0505219}
\begin{document}

\section{Introduction}
\label{sec:intro}

The `MHV rules' approach proposed in Ref.~\cite{CSW1}, has led to the 
establishment of
a new and powerful framework for computing large classes of
previously unknown tree-level and one-loop
scattering amplitudes in gauge theories, in a compact form, and
without appealing to Feynman diagrams.

In this paper, we apply the MHV rules to study the singular limits of QCD
amplitudes when $n$ partons (gluons and massless quarks) are simultaneously
collinear. This continues the program started in our earlier
work~\cite{Birthwright:2005ak} where MHV rules were used to derive
multi-collinear limits of amplitudes involving only gluons. Understanding the
infrared singular behaviour of tree-level QCD amplitudes is a prerequisite for
computing infrared-finite cross sections at fixed order in perturbation
theory.  In general, when one or more final state particles are either soft or
collinear, the amplitudes factorise. The first factor in this product is a
scattering amplitude that depends only on the remaining hard partons in the
process (including any hard partons constructed from an ensemble of unresolved
partons). The second factor is the splitting amplitude, it contains all of the
singularities due to the unresolved particles.   One of the best known examples
of this type of factorisation is the limit of tree amplitudes when two
particles are collinear.  This factorisation is universal and can be
generalised to more particles~\cite{Gehrmann-DeRidder:dblunres,Campbell:dblunres,Catani:NNLOcollfact,
Catani:IRtreeNNLO,delduca} and any number of
loops~\cite{Kosower:allorderfact}.

One of the main points of our approach \cite{Birthwright:2005ak}
is that, in order to derive all required splitting
functions we do not need to know the full amplitude. Out of the complete set of
MHV-diagrams contributing to the full amplitude, only a subset will contribute
in the multi-collinear limit. This subset includes only those MHV-diagrams
where {\em all} of the internal propagators go on-shell in the multi-collinear 
limit.
Moreover, the functions multiplying these singular propagators
in the splitting amplitude
are constrained by the MHV rules to take a purely holomorphic form: they are
functions which depend only on the holomorphic spinor products,
$\langle i ~j \rangle$, of the right-handed (undotted) spinors 
and not on the anti-holomorphic ones $[i~j]$.
This points towards a simple twistor space picture for the multi-collinear limits,
in terms of a degree-one curve in twistor space.
The MHV rules approach also enables us to calculate infinite
sequences of splitting amplitudes -- with fixed numbers of negative helicity partons
and arbitrary numbers of positive helicity ones, or vice versa.

The basic building blocks
of the MHV rules approach~\cite{CSW1}
are the colour-ordered
$n$-point vertices  which are connected by scalar propagators. These MHV
vertices are off-shell continuations of the maximally helicity-violating (MHV)
$n$-gluon scattering amplitudes  of Parke and Taylor~\cite{ParkeTaylor,BG}.   They
contain precisely two negative helicity gluons.
Written in terms of spinor inner
products~\cite{SpinorHelicity}, they are composed entirely of the holomorphic
products $\spa{i}.{j}$,
rather than their anti-holomorphic partners $\spb{i}.{j}$,
\be
 A_n(1^+,\ldots,p^-,\ldots,q^-,\ldots,n^+) =
\frac{\spa{p}.{q}^4}{ \spa1.2 \spa2.3 \cdots \spa{n-1,}.{n} \spa{n}.{1} },
\label{MHV}
\ee
where we introduce the common notation
$\spa{p_i}.{p_j}=\spa{i}.{j}$ and $\spb{p_i}.{p_j}=\spb{i}.{j}$.
By connecting MHV vertices,  amplitudes involving more
negative helicity gluons can be built up.

The MHV rules for gluons~\cite{CSW1} have been extended to amplitudes with
fermions~\cite{GK}.  New compact results for tree-level
gauge-theory results for non-MHV amplitudes  involving arbitrary numbers of
gluons~\cite{Zhu,KosowerNMHV,BBK}, and fermions~\cite{GK,GGK,Wu2} have been
derived.  They have been applied to processes involving external Higgs
bosons~\cite{DGK,BGK} and electroweak bosons~\cite{BFKM}.
MHV rules have also been shown to work at one-loop level for
supersymmetric theories~\cite{BST}.
Building on the earlier work of Bern, Dixon, Dunbar and Kosower~\cite{BDDK1,BDDK2},
there has been a remarkable progress in computing cut-constructible multi-leg
loop amplitudes in $\Neqfour$~\cite{CSW2,BST,Cachazo,
BCF1,BDDK7,BCF3,BDKNMHV}
and $\Neqone$~\cite{QuigleyRozali,BBST1,BBDD,BBDP1,Britto:2005ha}
supersymmetric gauge theories.  Encouraging
progress has also been made using MHV rules for non-supersymmetric loop
amplitudes~\cite{BBST2,BBDP2}.

Remarkably, the expressions obtained for the infrared singular parts of
$\Neqfour$ one-loop amplitudes (which are known to be  proportional to
tree-level results) were found to produce even more compact expressions for
gluonic tree amplitudes~\cite{BDKNMHV,RSV}. This observation led to the BCF
recursion relations~\cite{BCF4,BCFW} of Britto, Cachazo, Feng and Witten
as well as extremely
compact six-parton amplitudes~\cite{BCF4,LW1,LW2}. These tree-level BCF recursion relations
for massless particles have recently been generalised in two ways.
In Refs.~\cite{BDKrec,BDKrec2} a new version of recursion relations was adopted to calculate
all finite one-loop amplitudes in non-supersymmetric QCD.
At the same time, Ref.~\cite{Badger:2005zh} generalised BCF recursion relations to
include massive particles at tree level.

A comprehensive list of references
and a more detailed discussion of recent developments
can be found in the recent review \cite{Cachazo:2005ga}.
This progress has been stimulated by the original proposal of Witten
in \cite{Witten1} of a weak-to-weak coupling duality between a perturbative
${\cal N}=4$ gauge theory and a topological string theory in twistor space.

\medskip

The factorisation properties of amplitudes in the infrared
play several roles in developing higher order
perturbative predictions for observable quantities. First, a detailed
knowledge of the structure of unresolved emission enables phase space
integrations to be organised such that the infrared singularities due to soft
or collinear emission can be analytically
subtracted at NLO~\cite{Giele:1992vf,Frixione:1996ms,Catani:1997vz}
or at NNLO~\cite{Gehrmann-DeRidder:antenna}.    Second, they
enable large logarithmic corrections to be identified and resummed.
Third, the collinear limit plays a crucial role in the unitarity-based method
for loop calculations~\cite{BDDK1,BDDK2,Bern:1996db,Bern:1996je}.

In general, to compute a cross section at N$^n$LO, one requires detailed
knowledge of the infrared factorisation functions describing the unresolved
configurations for $n$-particles at tree-level, $(n-1)$-particles at one-loop
etc. The universal behaviour in the double collinear limit is well known at
tree-level (see for example Refs.~\cite{Altarelli:1977zs,Bassetto:1983ik}),
one-loop~\cite{Bern:1995ix,BDDK1,Bern:split1gluon,
Kosower:split1,Bern:split1QCD,Catani:2000pi} and at
two-loops~\cite{Bern:2lsplit,Badger:2lsplit}. Similarly, the triple collinear
limit has been studied at
tree-level~\cite{Gehrmann-DeRidder:dblunres,Campbell:dblunres,Catani:NNLOcollfact,Catani:IRtreeNNLO} and,
in the case of distinct quarks, at one-loop~\cite{Catani:2003}. Finally, the
tree-level quadruple gluon collinear limit was derived in Ref.~\cite{delduca,Birthwright:2005ak}.

\medskip

Our paper is organised as follows.   In Section~{\bf \ref{sec:col}}, we
briefly review the colour ordered formalism that underpins the MHV rules. 
The relevant MHV vertices are given in Section~{\bf \ref{sec:mhv}}.
Section~{\bf \ref{sec:limit1}} describes the procedure for taking the
collinear limit while the analytic structure of the splitting functions is
discussed in Section~{\bf \ref{sec:limit2}}.   
We write down general collinear
factorization formulae in Section~{\bf \ref{sec:general-results}}, which
are valid for specific numbers of negative helicity partons and an
arbitrary number of positive helicity partons.  
These results involve quarks and gluons in the collinear set and are
complementary to the multi-gluon splitting functions derived in
Ref.~\cite{Birthwright:2005ak}.
Specific explicit results
for the collinear limits of up to three collinear partons
are given in Sec.~{\bf
\ref{sec:specific-results}}.   
Our findings are
summarized in Sec.~{\bf \ref{sec:conclusion}}.

\section{Colour-ordered amplitudes}
\label{sec:col}

Tree-level multi-particle amplitudes can be decomposed into
colour-ordered partial amplitudes.
For gluons only, this decomposition is given by
\begin{equation}
{\cal A}_n(\{p_i,\lambda_i,a_i\}) =
i g^{n-2}
\sum_{\sigma \in S_n/Z_n} {\rm Tr}(T^{a_{\sigma(1)}}\cdots T^{a_{\sigma(n)}})\,
A_n(\sigma(1^{\lambda_1},\ldots,n^{\lambda_n}))\,.
\label{TreeColourDecomposition}
\end{equation}
Here $S_n/Z_n$ is the group of non-cyclic permutations on $n$
symbols, and $j^{\lambda_j}$ labels the momentum $p_j$ and helicity
$\lambda_j$ of the $j^{\rm th}$ gluon, which carries the adjoint
representation index $a_i$.  The $T^{a_i}$ are fundamental
representation SU$(N_c)$ colour matrices, normalized so that
${\rm Tr}(T^a T^b) = \delta^{ab}$.  The strong coupling constant is
$\alpha_s=g^2/(4\pi)$.
Note that the MHV rules method of Ref.~\cite{CSW1} is used to evaluate only the
purely kinematic amplitudes $A_n.$
Full amplitudes are then determined uniquely from the kinematic part $A_n$,
and the known expressions for the colour traces.

For processes involving a quark-antiquark pair and an arbitrary number of
gluons, the colour
decomposition is given by
\begin{eqnarray}
\lefteqn{
{\cal A}_n(\{p_i,\lambda_i,a_i\},\{p_j,\lambda_j,i_j\}) }\\
&&=
i g^{n-2}
\sum_{\sigma \in S_{n-2}} (T^{a_{\sigma(2)}}\cdots T^{a_{\sigma(n-1)}})_{i_1i_n}\,
A_n(1_q^{\lambda_1},\sigma(2^{\lambda_2},\ldots,{(n-1)}^{\lambda_{n-1}}),
n_{\bar q}^{\lambda_n})\,,\nonumber
\label{TreeColorDecomposition}
\end{eqnarray}
where $S_{n-2}$ is the set of permutations of $(n-2)$ gluons
and the fermions carry the fundamental colour labels $i_1$ and $i_n$.
By current conservation, the quark and antiquark helicities are  related such
that $\lambda_1 = -\lambda_n \equiv \lambda$ where $\lambda = \pm \frac{1}{2}$.

When an additional photon with momentum $P_{\gamma}$ is emitted,
the amplitudes have the following form,
\begin{eqnarray}
\lefteqn{
{\cal A}_n(\{p_i,\lambda_i,a_i\},\{p_j,\lambda_j,i_j\},P_\gamma) }\\
&&=
i e g^{n-2}
\sum_{\sigma \in S_{n-2}} (T^{a_{\sigma(2)}}\cdots T^{a_{\sigma(n-1)}})_{i_1i_n}\,
\tilde{A}_n(1_q^{\lambda_1},\sigma(2^{\lambda_2},\ldots,{(n-1)}^{\lambda_{n-1}}),
n_{\bar q}^{\lambda_n};P_{\gamma})\,,\nonumber
\label{qqgammacoldecomp}
\end{eqnarray}
where $e$ is the electric charge of the quark.

When there are two quark-antiquark pairs the
tree-level amplitude can be decomposed into colour ordered
amplitudes as,
\begin{eqnarray}
    &&\mathcal{A}_n(\{p_i,\lambda_i,a_i\},\{p_j,\lambda_j,i_j\})
    = i g^{n-2} \sum_k^{n-4}\sum_{\sigma\in S_k}\sum_{\rho\in S_l} \bigg\{ \nonumber\\
    &\phantom{-}&(T^{a_{\sigma(1)}}\cdots T^{a_{\sigma(k)}})_{i_1 i_n}
    (T^{a_{\rho(1)}}\cdots T^{a_{\rho(l)}})_{i_{s+1} i_{s}} \nonumber\\
    &&\times
    A_n(1_q^{\lambda},\sigma(1),\ldots,\sigma(k)),s^{-\lambda^\prime}_{\bar{Q}};
    (s+1)^{\lambda^\prime}_{Q},\rho(1),\ldots,\rho(l),n^{-\lambda}_{\bar{q}}) \nonumber\\
    &-&\frac{1}{N}(T^{a_{\sigma(1)}}\cdots T^{a_{\sigma(k)}})_{i_1 i_s} (T^{a_{\rho(1)}}\cdots T^{a_{\rho(l)}})_{
    i_{s+1}i_n} \nonumber\\
    &&\times\tilde{A}_n(1_q^{\lambda},\sigma(1),\ldots,\sigma(k),s^{-\lambda}_{\bar{q}};
    (s+1)^{\lambda^\prime}_{Q},\rho(1),\ldots,\rho(l),n_{\bar Q}^{-\lambda^\prime})\bigg\}
    \label{eq:qqQQcoldecomp}
\end{eqnarray}
where $S_k$ and $S_l$ are permutation groups such that $k+l=n-4$ and
represent the possible ways of distributing the gluons in a colour ordered way between
the quarks. For $i=j=0$, $(T^{a_i}\ldots T^{a_j})_{kl}$ reduces to $\delta_{kl}$.
We see that the two amplitudes $A_n$ and $\tilde{A}_n$ correspond to
different ways of connecting the fundamental colour charges.
For the $A$ amplitudes, there is a colour line connecting $q$ and $\bar Q$ and a second
line connecting $Q$ and $\bar q$, while for
the QED-like $\tilde A$ amplitudes the colour lines connect $q$ to $\bar q$ and $Q$ to
$\bar Q$. Any number of gluons may be radiated from each colour
line.
As before, by current conservation, the quark and antiquark helicities are  related such
that $\lambda_q = -\lambda_{\bar q} \equiv \lambda$
and $\lambda_Q = -\lambda_{\bar Q} \equiv \lambda^\prime$
where $\lambda, ~\lambda^\prime = \pm \frac{1}{2}$.

\section{MHV amplitudes}
\label{sec:mhv}

The colour ordered $n$-gluon MHV amplitude is given by
\begin{equation}
  \label{eq:gluon}
  A_n(1^+,\ldots,m_1^-,\ldots,m_2^-,\ldots,n^+) = \frac{\aab{m_1}{m_2}^4}{\prod_{i=1}^{n}\aab{i}{i+1}},
\end{equation}
while the two-quark multi-gluon  MHV amplitudes are,
\begin{eqnarray}
    A_n(1^{\lambda}_q,\ldots,m^-,\ldots,n^{-\lambda}_{\bar{q}})
    &=& \frac{\spa{m}.{1}^{2-2\lambda}\spa{m}.{n}^{2+2\lambda}}{\prod_{l=1}^n\spa{l}.{l+1}}
    \label{eq:2qMHV}.
\end{eqnarray}
Here the helicity of the quark is denoted by $\lambda = \pm \frac{1}{2}$ while $\ldots$ denotes an arbitrary number
of positive helicity gluons.
Amplitudes for a quark-antiquark pair, many gluons and a photon are given by,
\begin{eqnarray}
    \label{eq:2qMHVphoton}
\tilde{A}( 1_q^\lambda,\ldots,n_{\bar q}^{-\lambda};P_{\gamma}^-)
&=& \frac{\spa{P}.{1}^{2-2\lambda} \spa{P}.{n}^{2+2\lambda}}
{\spa{P}.{1}\spa{1}.{2}\cdots \spa{n}.{P}}, \\
\tilde{A}(1_q^\lambda,\ldots,m^-,\ldots,n_{\bar q}^{-\lambda};P_{\gamma}^+) &=&
\frac{\spa{m}.{1}^{2-2\lambda} \spa{m}.{n}^{2+2\lambda}}
{\spa{P}.{1}\spa{1}.{2}\cdots \spa{n}.{P}}.
\end{eqnarray}

In the four-quark case, there are four MHV amplitudes where two of the fermions have
negative helicity and two have positive helicity for each colour structure. For each
helicity configuration we can write,
\begin{eqnarray}
\label{eq:4qmhv1}
    A_n(1^+_q,\ldots,s^-_{\bar{Q}},(s+1)^+_{Q},\ldots,n^-_{\bar{q}}) &=& \frac{\spa{1}.{s}\spa{s}.{n}^2\spa{n}.{s+1}}{\prod_{l=1}^n \spa{l}.{l+1}} ,\\
\label{eq:4qmhv2}
    A_n(1^+_q,\ldots,s^+_{\bar{Q}},(s+1)^-_{Q},\ldots,n^-_{\bar{q}}) &=& \frac{\spa{1}.{s}\spa{n}.{s+1}^3}{\prod_{l=1}^n \spa{l}.{l+1}} ,\\
\label{eq:4qmhv3}
    A_n(1^-_q,\ldots,s^+_{\bar{Q}},(s+1)^-_{Q},\ldots,n^+_{\bar{q}}) &=&
    \frac{\spa{1}.{s}\spa{1}.{s+1}^2\spa{n}.{s+1}}{\prod_{l=1}^n \spa{l}.{l+1}} ,\\
\label{eq:4qmhv4}
    A_n(1^-_q,\ldots,s^-_{\bar{Q}},(s+1)^+_{Q},\ldots,n^+_{\bar{q}}) &=& \frac{\spa{1}.{s}^3\spa{n}.{s+1}}{\prod_{l=1}^n \spa{l}.{l+1}},
\end{eqnarray}
with the other colour ordering given by,
\begin{eqnarray}
\label{eq:4qmhv5}
    \tilde{A}_n(1^+_q,\ldots,s^-_{\bar{q}},(s+1)^+_{Q},\ldots,n^-_{\bar{Q}}) &=& \frac{\spa{1}.{n}\spa{n}.{s}^2\spa{s}.{s+1}}{\prod_{l=1}^n \spa{l}.{l+1}} ,\\
\label{eq:4qmhv6}
    \tilde{A}_n(1^+_q,\ldots,s^-_{\bar{q}},(s+1)^-_{Q},\ldots,n^+_{\bar{Q}}) &=& \frac{\spa{1}.{n}\spa{s}.{s+1}^3}{\prod_{l=1}^n \spa{l}.{l+1}} ,\\
\label{eq:4qmhv7}
    \tilde{A}_n(1^-_q,\ldots,s^+_{\bar{q}},(s+1)^-_{Q},\ldots,n^+_{\bar{Q}})
    &=& \frac{\spa{1}.{n}\spa{1}.{s+1}^2\spa{s}.{s+1}}{\prod_{l=1}^n \spa{l}.{l+1}} ,\\
\label{eq:4qmhv8}
    \tilde{A}_n(1^-_q,\ldots,s^+_{\bar{q}},(s+1)^+_{Q},\ldots,n^-_{\bar{Q}}) &=& \frac{\spa{1}.{n}^3\spa{s}.{s+1}}{\prod_{l=1}^n \spa{l}.{l+1}}.
    \label{eq:4qMHVamps2}
\end{eqnarray}

The  \MHVb~amplitudes are related by parity and can be obtained by conjugating
the MHV expressions,
\begin{equation}
    A_n(1^{\lambda_1},\ldots,n^{\lambda_n}) =
    (-1)^n\left(A_n(1^{-\lambda_1},\ldots,n^{-\lambda_n})\right)^*,
\end{equation}
and similarly for the $\tilde{A}$ amplitudes.

\section{Collinear limits}
\label{sec:limit1}

To find the splitting functions we work with the colour stripped amplitudes.
For these colour ordered amplitudes, it is known  that when the
collinear particles are not adjacent there is no collinear
divergence~\cite{delduca}. Therefore, without loss of generality,  we can take
particles $1 \dots n$ collinear.

The multiple collinear limit is approached when the momenta $p_1,
\dots, p_n$ become parallel.  This implies that all the
particle subenergies $s_{ij}=(p_i+p_j)^2$, with $i,j=1,\dots,n$, are
simultaneously small. We thus introduce a pair of
light-like momenta $P^\nu$
and $\xi^\nu$ ($P^2=0, \xi^2=0$), and we write
\begin{equation}
(p_1 + \dots + p_n)^\nu = P^\nu
+ \frac{s_{1,n} \; \xi^\nu}{2 \, \xi \cdot P} \;, \quad
s_{i,j} = (p_i + \dots + p_j)^2 \;,
\end{equation}
where $s_{1,n}$ is the total invariant mass of the system of collinear
partons. In the collinear limit, the vector $P^\nu$
denotes the collinear direction, and the individual collinear momenta are
$p_i^\nu \to z_i P^\nu$. Here the longitudinal-momentum
fractions $z_i$ are given by
\begin{equation}
z_i = \frac{\xi \cdot p_i}{\xi \cdot P}
\end{equation}
and fulfil the constraint $\sum_{i=1}^m z_i =1$.
To be definite, in the rest of the paper we work
in the time-like region so that
($s_{ij} > 0, \;  1>z_i > 0$).

\begin{figure}[htbp]
  \centering
  \psfrag{-P-l}{$-P^{-\lambda}$}
  \psfrag{Pl}{$P^{\lambda}$}
  \psfrag{pn+1}{$(n+1)$}
  \psfrag{pn}{$n$}
  \psfrag{pN}{$N$}
  \psfrag{p1}{$1$}
  \includegraphics[width=12cm]{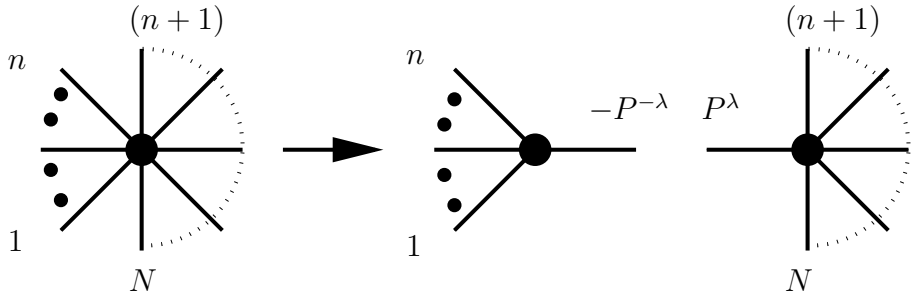}
  \caption{Factorisation of an $N$-point colour ordered amplitude with gluons $p_1,\ldots,p_n$ collinear
  into splitting function for $P \to 1, \ldots, n$
  multiplied by an $(N-n+1)$-point amplitude.}
  \label{fig:limit}
\end{figure}

As illustrated in Fig.~\ref{fig:limit},
in the multi-collinear limit an $N$-particle colour ordered tree amplitude
factorises
and can be written as
\begin{eqnarray}
  \label{eq:factorise}
  A_N(1^{\lambda_1},\ldots,N^{\lambda_N}) &\to&
  \ssplit({{1}^{\lambda_{1}},\ldots,n^{\lambda_n} \to P^\lambda})
  \times
  A_{N-n+1}((n+1)^{\lambda_{n+1}},\ldots ,N^{\lambda_N},P^\lambda).\nonumber \\
\end{eqnarray}
This labelling of the splitting amplitude
$\ssplit({1^{\lambda_1},\ldots,n^{\lambda_n}\to P^\lambda})$ differs from the
usual definition because we use the momentum and helicity that
participates in the resultant amplitude $P^\lambda$ rather than $-P^{-\lambda}$.
With this choice, it is easier to see how the helicity is conserved
in the splitting, i.e.   helicity $\lambda^1,\ldots,\lambda^n$ is replaced by
$\lambda$.   Since eq.~(\ref{eq:factorise}) applies for all $N$, we can
use it to derive the splitting amplitude by systematically choosing $N = 3+n$.
In this case, we always factorise onto a four-point amplitude.

\section{Analytic structure of splitting amplitudes}
\label{sec:limit2}

The MHV rules of Ref.~\cite{CSW1}
were developed for calculating
purely gluonic amplitudes at tree level and extended to
amplitudes involving fermions in Ref.~\cite{GK}. In this approach
all non-MHV $N$-particle
amplitudes (including $\overline{\rm MHV}$) are expressed
as sums of tree diagrams in an effective scalar perturbation theory.
The vertices in this theory are the MHV amplitudes of Eq.~(\ref{MHV})
continued off-shell and connected by scalar
propagators $1/q^2$.

Following~\cite{Birthwright:2005ak}, we classify
collinear limits according to the difference between the number of negative
helicity particles before taking the collinear limit, and the number after,
$\Delta M$.
Splitting amplitudes are calculated using the factorisation formula
eq.~(\ref{eq:factorise}). To facilitate the calculation, it makes sense to factorise
onto hard amplitudes with the simplest analytic structure.
Hence, in the MHV-rules formalism
we will always factorise
onto MHV amplitudes which are listed in section \ref{sec:mhv}.
In this case we find that $\Delta M$ of the splitting amplitude
satisfies the relation,
\be
\Delta M + 2\, =\, N_{-}
\ee
where $2$ is the number of negative helicities in the hard MHV amplitude, and
$N_{-}$ is the total number of negative helicities in the full amplitude.
$\Delta M$ determines the
order of MHV diagram~\cite{CSW1} for the full amplitude $A_N$
\bea
\Delta M=0 \quad \Rightarrow  \quad
&&1^+,2^+,3^+, \ldots ,n^+  \rightarrow P^+ \nonumber \qquad A_N={\rm MHV}\\
&&1^-,2^+,3^+, \ldots ,n^+  \rightarrow P^-\nonumber\\
    & &\nonumber\\
\Delta M=1 \quad \Rightarrow \quad
&&1^-,2^+,3^+, \ldots ,n^+  \rightarrow P^+ \nonumber  \qquad A_N={\rm NMHV}\\
&&1^-,2^-,3^+, \ldots ,n^+  \rightarrow P^- \label{mmp} \nonumber\\
  & &\nonumber\\
\Delta M=2 \quad \Rightarrow  \quad
&&1^-,2^-,3^+, \ldots ,n^+  \rightarrow P^+\nonumber  \qquad A_N={\rm NNMHV}\\
&&1^-,2^-,3^-, \ldots ,n^+  \rightarrow P^-\nonumber\\
\eea
and so on for all $\Delta M>2$ cases.

If we choose to use \MHVb\ rules, we extract the splitting function
by factorising onto \MHVb\ amplitudes.
Splitting amplitudes are then classified by
the difference in the number of positive helicity particles,
$\Delta P$, and similar observations apply.

In general, any splitting amplitude can be obtained from either MHV or
 \MHVb\ rules.
A simple power counting argument~\cite{Birthwright:2005ak} gives
\be
\label{powerct}
\ssplit \, \propto \,
\frac{1}{\spb{\ }.{\ }^{\Delta M} \spa{\ }.{\ }^{\Delta P}} \ .
\ee
For an MHV-rules diagram to contribute to $ \Delta M \neq 0$ collinear limits,
it
must contain anti-holomorphic spinor products $\spb{i}.{j}$ of
collinear momenta. However, because on-shell MHV vertices are
entirely holomorphic,
within the MHV rules there are only two potential sources of the
anti-holomorphic spinor products. One source is scalar propagators
$1/s_{ij}=1/\spa{i}.{j} \spb{j}.{i}$ which connect MHV vertices. The
second source is the off-shell continuation of the corresponding connected
legs in the MHV vertices.  Each off-shell continued leg of momentum $P$ gives
rise to a factor $\langle i  P \rangle \propto \langle i | P | \eta]$ which
amounts to anti-holomorphic factors of the form $[j \eta]$.
When the reference
spinors $\eta_{\dot\alpha}$ are kept general, the
$\eta$-dependence must cancel and therefore
the off-shell continuation cannot give rise
to an overall factor of $\spb{i}.{j}$.

This implies that within the MHV rules, the anti-holomorphic
spinor products
in \eqref{powerct} arise solely from the internal propagators.
Since $\Delta M= v_{MHV}-1$,\footnote{In principle,
$\Delta M= v_{MHV}-v_{MHV}^\prime$ where $v_{MHV}^\prime$ is the number of
MHV vertices remaining in the factored amplitude.   The splitting function is
independent of $v_{MHV}^\prime$, and because we systematically choose
to factor directly onto a single MHV vertex, we set $v_{MHV}^\prime = 1$.}
 where $v_{MHV}$ is the number of MHV vertices
in the diagram, the total number of internal propagators is $\Delta M$,
in agreement with \eqref{powerct}.
Similarly, in the \MHVb\ approach, the holomorphic products would arise solely
from internal propagators whose total number in \MHVb\ diagrams is
$\Delta P$.

More precisely, it follows that all
splitting amplitudes can be recast as
\bea
\ssplit & = & \sum \, \frac{1}{\prod_{{i,j}=1}^{\Delta M} s_{i,j}} \,\, f(\spa{ }.{ })
\label{54}\\
& = & \sum \, \frac{1}{\prod_{{i,j}=1}^{\Delta P} s_{i,j}} \,\, \tilde{f}(\spb{}.{})
\label{55}
\eea
where the first expression follows from the MHV rules representation,
and the second expression -- from the \MHVb\ formalism.
Here the summations are over all inequivalent choices of $\Delta M$
($\Delta P $) products of vanishing kinematic
invariants $s_{i,j}$ which corresponds to
different MHV (\MHVb) rules diagrams.
The
coefficient functions $f$ depend only on holomorphic spinor products, while
the \MHVb\ coefficients
$\tilde{f}$ are purely anti-holomorphic. Moreover, $f$ and $\tilde f$ have
dimensions,
\be
f \, \propto\, \frac{1}{\spa{\ }.{\ }^{\Delta P -\Delta M}} \ , \qquad
\tilde{f} \, \propto\, \frac{1}{\spb{\ }.{\ }^{\Delta M -\Delta P}} \ .
\ee
The fact that $f$ ($\tilde{f}$) is purely (anti)-holomorphic suggests a simple
twistor-space interpretation. All splitting functions can be represented as
sums over the corresponding poles in $s$ with the coefficients being supported
on a single degree-one curve in (anti)-twistor space. This pure (anti)-holomorphic
representation of multi-collinear limits is specific to the MHV (\MHVb) formalism
and is lost in the usual Feynman-diagram-type approaches as in Ref.~\cite{delduca},
or in the BCF recursive
approach, as shown in \cite{Birthwright:2005ak}.

We further note that MHV rules for collinear limits are substantially
simpler than the rules
for the full amplitudes. Collinear splitting functions follow from a
{\it subset} of the MHV rules diagrams
\cite{Birthwright:2005ak}. The subset is determined by requiring that all internal
propagators are on-shell in the multi-collinear limit.\footnote{This is dictated
by eq.~\eqref{54} in the MHV formalism (or eq.~\eqref{55} for \MHVb\ rules).}
This is a powerful
constraint on the types of the contributing diagrams and it simplifies
taking the collinear limit dramatically.

As mentioned earlier, each splitting amplitude can be calculated in both
the MHV and
in the \MHVb\ approaches.
In practice, eqs.~\eqref{powerct}-\eqref{55} imply that
the MHV approach is simpler if $\Delta M < \Delta P$,
while the \MHVb\ approach is more
compact in the opposite case, $\Delta P < \Delta M$.

In most of what follows we will concentrate on the splitting amplitudes
with $\Delta M \le \Delta P$ and will follow the MHV rules. The remaining
amplitudes with $\Delta P < \Delta M$ are obtained from
these by complex conjugation.

\subsection{ An example}

When $\Delta M = \Delta P$ both MHV and \MHVb\ rules are expected
to yield results of similar complexity.
As an example, let us consider a triple collinear splitting with $\Delta M = \Delta P = 1$.
In full generality, the MHV (\MHVb) rules approach should generate a maximum of three terms
corresponding to simple poles in $s_{1,2}$, $s_{2,3}$ and $s_{1,3}\equiv (p_1+p_2+p_3)^2$.
For the specific splitting
$1_q^-,2_{\bar Q}^+,3_Q^-\to P_q^-$, the MHV rules approach yields,
\begin{eqnarray}
\label{eq:qpQpQm1}
  \ssplit(1_q^+,2_{\bar Q}^+,3_Q^- \to P_q^+)&=&
-{\frac {\spa{2}.{3}z_{{2}}}{s_{{2,3}} \left( \spa{1}.{2}\sqrt {z_{{2}
}}+\spa{1}.{3}\sqrt {z_{{3}}} \right)  \left( z_{{2}}+z_{{3}} \right)
}}\nonumber \\&&+{\frac { \left( \spa{1}.{3}\sqrt {z_{{1}}}+\spa{2}.{3}\sqrt {z_{{2}
}} \right) ^{2}}{s_{{1,3}} \left( \spa{1}.{2}\sqrt {z_{{2}}}+\spa{1}.{
3}\sqrt {z_{{3}}} \right) \spa{2}.{3}}}\ ,
\end{eqnarray}
while the \MHVb\ rules approach finds,
\begin{eqnarray}
\label{eq:qpQpQm2}
  \ssplit(1_q^+,2_{\bar Q}^+,3_Q^- \to P_q^+)&=&
{\frac {z_{{1}}z_{{3}}\spb{2}.{3}}{s_{{2,3}}
\left( \spb{1}.{2}\sqrt {z_{{2}}}+\spb{1}.{3}\sqrt {z_{{3}}} \right)
\left( z_{{2}}+z_{{3}} \right)
}}\nonumber \\&&
-{\frac { \spb{1}.{2}^2}{s_{{1,3}} \left( \spb{1}.{2}\sqrt {z_{{2}}}+\spb{1}.{
3}\sqrt {z_{{3}}} \right) \spb{2}.{3}}}\ .
\end{eqnarray}
As expected, the $s_{12}$ pole is absent because there is no $q \bar Q$ collinear limit.
By taking the limit of a Feynman diagram calculation, Ref.~\cite{delduca}
finds,
\begin{eqnarray}
\label{eq:qpQpQm3}
  \ssplit(1_q^+,2_{\bar Q}^+,3_Q^- \to P_q^+)&=&
\frac {\sqrt{z_1z_2z_3}}{s_{2,3}(z_2+z_3)}
+\frac { \spb{1}.{2}
\left( \spa{1}.{3}\sqrt {z_1}+\spa{2}.{3}\sqrt {z_2} \right)}{s_{1,3}s_{2,3}}\ .
\end{eqnarray}
Results \eqref{eq:qpQpQm1}, \eqref{eq:qpQpQm2} and \eqref{eq:qpQpQm3}
are for the same amplitude and
all three expressions agree numerically. But the analytic form of these specific
representations is different.
In agreement with eqs.~\eqref{54}-\eqref{55},
the functions accompanying the $1/s$ poles,  are holomorphic in the
MHV result \eqref{eq:qpQpQm1},
are anti-holomorphic in the \MHVb\ expression \eqref{eq:qpQpQm2},
while  the Feynman diagram result \eqref{eq:qpQpQm3} contains a
mixture of holomorphic and anti-holomorphic terms.
(In this case, it happens to give a more compact result.)
In general, the limit of an amplitude computed using the BCF recursion relations
will also provide a mixed holomorphic/anti-holomorphic splitting function
(as discussed in sections {\bf 4.2.2} and {\bf 4.2.3} of Ref.~\cite{Birthwright:2005ak}).
In this specific case, taking the collinear limit of the compact
expression for the appropriate six-parton amplitude given in Ref.~\cite{LW2} exactly
reproduces the MHV result of eq.~\eqref{eq:qpQpQm2}.

\section{General results}
\label{sec:general-results}

In this section we give the results for the multiple collinear limit
of quarks and gluons.
We categorise the results according to the number of quarks
involved in the limit.
In each case,  we give the general results for
collinear limits with $\Delta M =\, 0\, , \, 1$ and involving
an arbitrary number of positive helicity particles.

Limits of the type
$\ssplit({1^+,\ldots,n^+\to P^+})$ and
$\ssplit({1^-,2^+,\ldots,n^+ \to P^-})$ can contribute to the $\Delta M=0$,
and these collinear splitting functions
are straightforward to derive directly from the simple MHV vertex.

For the remaining splitting functions, it is useful
to introduce the more compact notation
\begin{eqnarray}
\ssplit(1^+,\ldots , m_1^-, \ldots, m_2^-, \ldots,m_r^-,\ldots ,
  n^+ \to P^\pm)=\Split_{\pm}(m_1,\ldots,m_r)\ .
\end{eqnarray}

\begin{figure}[t!]
    \centering
    \psfrag{i+1+}{$~$}
    \psfrag{a}{$(a)$}
    \psfrag{b}{$(b)$}
    \psfrag{j+1+}{$~$}
    \psfrag{j+}{$j+$}
    \psfrag{i+}{{\small $i+$}}
    \psfrag{m1-}{$m_1^-$}
    \psfrag{m2-}{$m_2^-$}
    \begin{center}
        \includegraphics[width=12cm]{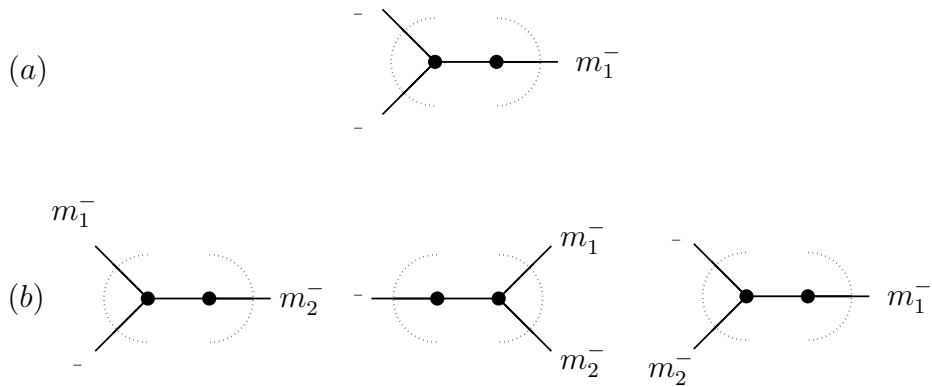}
    \end{center}
    \caption{MHV topologies contributing to (a) $\Split_+(m_1)$ and (b) $\Split_-(m_1,m_2)$.  Negative helicity particles are indicated
    by solid lines, while
    arbitrary numbers of positive helicity particles emitted from each vertex are shown as dotted arcs.
    All particles that are not in the collinear set must be emitted from the left-hand vertex.}
\label{fig:Split}
\end{figure}

For $\Delta M = 1$, there are two possible types of splitting function,
$\Split_{+}(m_1)$ and $\Split_{-}(m_1,\ldots,m_r)$.
The possible MHV topologies contributing to these splitting functions are illustrated in
Fig.~\ref{fig:Split}.  Only negative helicity particles are
shown. In the collinear limit, the propagator goes on-shell.
Any MHV diagram with a hard particles emitted from both vertices produces an off-shell
propagator.
This means that {\em only} particles from the collinear set are allowed to
couple to the right-hand vertex.  All hard partons couple to the left-hand vertex.

Throughout we adopt the notation of Ref.~\cite{Birthwright:2005ak}.
In order that the limits can be read directly from the MHV diagrams,
we make the following substitutions.
If $a$ is a
particle from the collinear set,
$b$ is a particle which is not in the collinear set, and
$q$ is the sum of the collinear momenta from $i+1$ to $j$, then
\bea
  \spa{a}.{q}&\rightarrow&\, \spb{P}.{\eta}\sum_{l=i+1}^{j}\spa{a}.{l}\sqrt{z_l} \, \equiv\,
  \spb{P}.{\eta}\Del{i}.{j}.{a}, \label{aqqq}\\
  \spa{b}.{q}&\rightarrow&\, \spb{P}.{\eta}\spa{b}.{P}\sum_{l=i+1}^{j} z_l ,\label{Xq}\\
  \spa{b}.{a}&\rightarrow&\, \spa{b}.{P} \sqrt{z_a} \ .\label{Xa}
\eea
The $\Delta$ is defined as
\begin{equation}
  \label{eq:7}
  \Del{i}.{j}.{a} = \sum_{l=i+1}^{j}\spa{a}.{l}\sqrt{z_l} \ ,
\end{equation}
noting that
the
 boundary terms involving either $\spa{0}.{1}$ or $\spa{n}.{n+1}$,
 are given by,
\begin{eqnarray}
 \frac{\spa{n}.{n+1}}{\Del{i}.{n}.{n+1}} &\rightarrow&\,
 -\frac{\sqrt{z_n}}{\sum_{l=i+1}^{n}z_l}, \label{jn}\\
 \frac{\spa{0}.{1}}{\Del{0}.{j}.{0}} &\rightarrow&\,
 \frac{\sqrt{z_1}}{\sum_{k=1}^{j}z_k}.\label{inp3}
\end{eqnarray}
We  also introduce
\begin{eqnarray}
\label{D}
D(i,j,q_{i+1,j})&=&\frac{q_{i+1,j}^2}{\spa{i,}.{i+1} \spa{j,}.{j+1}}
\Del{i}.{j}.{i}\Del{i}.{j}.{i+1}\Del{i}.{j}.{j}\Del{i}.{j}.{j+1}\ .\nonumber \\
\end{eqnarray}

\subsection{One quark in the collinear set: $q(ng) \to q$}

\subsubsection{$\Delta M = 0$}

This is the simplest case which is read directly off the single MHV vertex.
For positive helicity quarks, we
use the two-quark MHV amplitude of Eq.~(\ref{eq:2qMHV}) and find,
\begin{eqnarray}
\ssplit(1_q^+ ,\ldots, n^+ \to P_q^+)= \frac{\sqrt{z_1}}{\sqrt{z_1 z_n} \prod_{l=1}^{n-1} \spa{l,}.{l+1} } \ .
\end{eqnarray}
For negative helicity quarks,
\begin{eqnarray}
\ssplit(1_q^- ,\ldots, n^+ \to P_q^-)= \frac{\sqrt{z_1}^3}{\sqrt{z_1 z_n} \prod_{l=1}^{n-1} \spa{l,}.{l+1} } \ .
\end{eqnarray}
Note that helicity conservation ensures that the helicity of $P$ is the same as that of $q$.
It is often convenient to combine results for quarks of helicity $\lambda = \pm
\frac{1}{2}$ such that,
\begin{eqnarray}
\ssplit(1_q^\lambda ,\ldots, n^+ \to P_q^\lambda)=
\frac{\sqrt{z_1}^{2-2\lambda}}{\sqrt{z_1 z_n} \prod_{l=1}^{n-1} \spa{l,}.{l+1} } \ .
\end{eqnarray}

Using parity we find,
\begin{eqnarray}
\ssplit(1_q^+ ,\ldots, n^- \to P_q^+)= \frac{(-1)^{n-1}\sqrt{z_1}^{2+2\lambda}}{\sqrt{z_1 z_n} \prod_{l=1}^{n-1} \spb{l,}.{l+1} } \ .
\end{eqnarray}

The amplitudes where an antiquark is collinear with several gluons are obtained by charge conjugation.

\subsubsection{$\Delta M = 1$}

Because of helicity conservation, $\Delta M=1$ implies that a single gluon has negative helicity.
When the quark has positive helicity, then the
MHV diagrams contributing in the collinear limit correspond to topology (a) of Fig.~\ref{fig:Split}.
There are two types of diagram -- one class where the quark is emitted from the right-hand vertex
(and the propagating particle
is a quark) and one class mediated by gluon exchange where the quark is emitted from the left-hand vertex.
We find,
\begin{eqnarray}
\label{eq:qgg1}
\lefteqn{  \ssplit(1_q^+,\ldots,m^-,\ldots,n^+\to P_q^+) =
  \frac{1}{\sqrt{z_1 z_n} \prod_{l=1}^{n-1} \spa{l,}.{l+1} }}\nonumber\\
 &&\hspace{2cm}\times\bigg[
  -\sum_{j=m}^{n}\frac{\Delta^3(0,j;m)
  \aab{1}{m}}{D(0,j,q_{1,j})}\left (\sum_{k=1}^{j}z_k\right ) +  \sum_{i=1}^{m-1} \sum_{j=m}^{n}
\frac{\Delta^4(i,j;m)}{D(i,j,q_{i+1,j})}\sqrt{z_1} \bigg].\nonumber \\
\end{eqnarray}
In the same manner,
for negative helicity quarks, the allowed MHV diagrams correspond to the first and second
topologies shown in  Fig.~\ref{fig:Split}(b),
\begin{eqnarray}
\label{eq:qgg2}
\lefteqn{  \ssplit(1_q^-,\ldots,m^-,\ldots,n^+\to P_q^-)=
  \frac{1}{\sqrt{z_1 z_n} \prod_{l=1}^{n-1} \spa{l,}.{l+1} }}\nonumber \\
 &&\hspace{2cm}\times\bigg[
  -\sum_{j=m}^{n}\frac{\Delta(0,j;m)
    \aab{1}{m}^3}{D(0,j,q_{1,j})}\left (\sum_{k=1}^{j}z_k \right)^3+  \sum_{i=1}^{m-1} \sum_{j=m}^{n}
\frac{\Delta^4(i,j;m)}{D(i,j,q_{i+1,j})}\sqrt{z_1}^3 \bigg].\nonumber \\
\end{eqnarray}

\subsection{Two quarks in the collinear set: $(ng)\bar qq \to g$ }

In this collinear limit, the $\bar q q$ pair is in the adjoint representation
and effectively acts as a gluon.

\subsubsection{$\Delta M = 0$}

This is the simplest case which is read directly off the single MHV vertex.
Unlike the previous case, here we start with a two-quark MHV amplitude and factorise onto
a gluonic MHV amplitude.   Alternatively, we could start with a four-quark amplitude and
factorise onto a two-quark amplitude.
For quarks with helicity $\lambda = \pm\frac{1}{2}$, we find,
\begin{eqnarray}
\ssplit(1^+ ,\ldots,s^{-\lambda}_{\bar q},(s+1)^\lambda_q,\ldots, n^+ \to P^-)=
\frac{\sqrt{z_s}^{2+2\lambda}\sqrt{z_{s+1}}^{2-2\lambda}}{\sqrt{z_1 z_n} \prod_{l=1}^{n-1} \spa{l,}.{l+1} } \ .
\end{eqnarray}

\subsubsection{$\Delta M = 1$}

For amplitudes of the $\Split_+(m_1)$-type, we find
\begin{eqnarray}
\label{eq:gqq}
\lefteqn{
\ssplit(1^+ ,\ldots,s^{-\lambda}_{\bar q},(s+1)^\lambda_q,\ldots, n^+ \to P^+)= }\nonumber \\
&&
  \frac{1}{\sqrt{z_1 z_n} \prod_{l=1}^{n-1} \spa{l,}.{l+1} }
  \sum_{i=0}^{s-1}\sum_{j=s+1}^{n}
  \frac{\Delta^{2+2\lambda}(i,j;s) \Delta^{2-2\lambda}(i,j;s+1)}{D(i,j,q_{i+1,j})}  .
\end{eqnarray}

There are four diagrams contributing to splitting functions of
$\Split_-(m_1,m_2)$ type\footnote{Diagrams where both the negative helicity fermion and gluon couple to the
right-hand vertex in Fig.~\ref{fig:Split}(b) can be mediated by either fermion or gluon exchange.},
\begin{eqnarray}
  \lefteqn{
  \ssplit(1^+,\ldots, s^{-\lambda}_{\bar q},(s+1)^\lambda_q,\ldots, m^-,\ldots,n^+ \to P^-)=
  \frac{1}{\sqrt{z_1 z_n} \prod_{l=1}^{n-1} \spa{l,}.{l+1} }}\nonumber \\
  &&\hspace{5cm}\times\bigg[
  \sum_{i=0}^{s-1}\sum_{j=m}^{n}
  \frac{\aab{s}{m}^{2+2\lambda}\aab{s+1}{m}^{2-2\lambda}}{D(i,j,q_{i+1,j})}
  \left(\sum_{k=i+1}^{j}z_k\right)^4 \nonumber\\
&&\hspace{5cm}+
  \sum_{i=0}^{s-1}\sum_{j=s+1}^{m-1}\frac{\Delta^{2+2\lambda}(i,j;s)
  \Delta^{2-2\lambda}(i,j;s+1)}{D(i,j,q_{i+1,j})} z_m^2 \nonumber\\
&&\hspace{5cm}- \sum_{j=m}^{n}
\frac{\Delta^{2+2\lambda}(s,j;m)\aab{s+1}{m}^{2-2\lambda}}
{D(s,j,q_{s+1,j})}\sqrt{z_s}^{2+\lambda}
  \left(\sum_{k=s+1}^{j} z_k \right)^{2-2\lambda} \nonumber\\
&&\hspace{5cm}+ \sum_{i=s+1}^{m-1}\sum_{j=m}^{n} \frac{\Delta^4(i,j;m)}{D(i,j,q_{i+1,j})}
\sqrt{z_s}^{2+2\lambda}\sqrt{z_{s+1}}^{2-2\lambda} \bigg].
\end{eqnarray}
Splitting functions of the type
$\ssplit(1^+,\ldots, m^-, \ldots, s^{-\lambda}_{\bar q},(s+1)^\lambda_q,\ldots,n^+ \to P^-)$ are obtained
by line reversal. These results for the two-quark sector are sufficient to calculate all
splitting amplitudes for up to four partons.

\subsection{Two quarks in the collinear set: $q(ng)\bar q \to \gamma$}

In this collinear limit, the $q\ldots \bar q $ system
forms a colour singlet and effectively acts as a photon.

\begin{figure}[t!]
    \centering
    \psfrag{n+1}{$n+1$}
    \psfrag{n}{$n^+$}
    \psfrag{N}{$N$}
    \psfrag{1}{$1^-$}
    \psfrag{m}{$m^-$}
    \begin{center}
        \includegraphics[height=5cm]{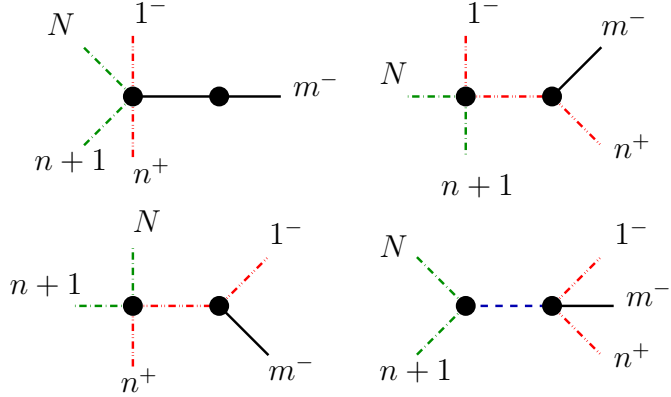}
    \end{center}
    \caption{MHV topologies contributing to the two
    quark collinear limit of the type
    $\widetilde{\ssplit}(1_q^-,\ldots,m^-,\ldots,n_{\bar q}^+\to P_\gamma^-)$.
    Quarks of type $Q$ ($q$) are shown as green(red)-dotdashed lines and negative
    helicity gluons as black solid lines. The negative helicity photon is shown
    as a blue dashed line.
 }
\label{fig:2qtilde}
\end{figure}

\subsubsection{$\Delta M = 0$}

In this limit the four-quark $\tilde A$ MHV amplitudes of
eqs.~(\ref{eq:4qmhv5})--(\ref{eq:4qmhv8}) factorise directly onto
the two-quark+photon amplitudes of (\ref{eq:2qMHVphoton}).
We find that,
\begin{eqnarray}
    \widetilde{\ssplit}(1_q^\lambda,\ldots,n_{\bar q}^{-\lambda}\to P_\gamma^-) &=&
   \frac{z_1^{\frac{1}{2}-\lambda}z_n^{\frac{1}{2}+\lambda}}
   {\spa{1}.{2}\cdots\spa{n-1}.{n}}.
\end{eqnarray}

\subsubsection{$\Delta M = 1$}

For amplitudes of the $\Split_+(m_1)$ type, there is a single MHV diagram and
we find
\begin{eqnarray}
\widetilde{\ssplit}(1_q^-,\ldots,n_{\bar q}^+\to P_\gamma^+) &=&
\frac{1}{\sqrt{z_1 z_n} \prod_{l=1}^{n-1} \spa{l,}.{l+1} }
  \times \frac{\Del{0}.{n}.{1}^3 \Del{0}.{n}.{n} }{D(0,n;q_{1,n})}. \nonumber
\\
\end{eqnarray}

As in the previous case, there are four diagrams shown in Fig.~\ref{fig:2qtilde}
contributing 
to splitting functions of
$\Split_-(m_1,m_2)$ type such that,
\begin{eqnarray}
\lefteqn{\widetilde{\ssplit}(1_q^-,\ldots,m^-,\ldots,n_{\bar q}^+\to P_\gamma^-) =
\frac{1}{\sqrt{z_1 z_n} \prod_{l=1}^{n-1} \spa{l,}.{l+1} }}\nonumber \\
  &&\hspace{2cm}\times\bigg[
\sum_{i=1}^{m-1} \sum_{j=m}^{n-1} \frac{\Del{i}.{j}.{m}^4
  \sqrt{z_1}^3\sqrt{z_n}}{D(i,j;q_{i+1,j})}-\sum_{i=1}^{m-1}
\frac{\spa{n}.{m}\Del{i}.{n}.{m}^3}{D(i,n;q_{i+1,n})}\sqrt{z_1}^3
\left(\sum_{k=i+1}^{n} z_k\right) \nonumber\\
 &&\hspace{2cm}-\sum_{j=m}^{n-1}
\frac{\spa{1}.{m}^3\Del{0}.{j}.{m}}{D(0,j;q_{1,j})}\sqrt{z_n}
\left(\sum_{k=1}^{j} z_k\right)^3 +
  \frac{\spa{1}.{m}^3\spa{n}.{m} }{D(0,n;q_{1,n})}
\bigg ].
\end{eqnarray}

\subsection{Three quarks in the collinear set: $q (ng) \bar Q Q \to q$}

In this configuration, the $\bar Q$ is adjacent with $Q$
and therefore the vertices in the MHV rules
include the
four-quark amplitudes of eqs.~(\ref{eq:4qmhv1})--(\ref{eq:4qmhv4}).
The factorised amplitude is a two-quark MHV as given in  eq.~(\ref{eq:2qMHV})
Furthermore, since the helicity of $q$ is conserved and the helicities
of $Q$ and $\bar Q$ are opposite, there are no $\Delta M=0$ splitting functions.

\subsubsection{$\Delta M = 1$}

For $\Delta M = 1$, the two diagrams (with quark and gluons exchanged)
of $\Split_+(m_1)$-type yield,
\begin{eqnarray}
\lefteqn{
 \ssplit(1_q^+,\ldots, s^{-}_{\bar Q},(s+1)^+_Q,\ldots,n^+ \to P_q^+)=
  \frac{1}{\sqrt{z_1 z_n} \prod_{l=1}^{n-1} \spa{l,}.{l+1}
  }\hspace{2cm}\phantom{~}.}\nonumber \\
  &&\hspace{4cm}\times\bigg[
 \sum_{i=1}^{s-1} \sum_{j=s+1}^n
\frac{\Delta(i,j;s+1)\Delta^3(i,j;s)}{D(i,j,q_{i+1,j})} \sqrt{z_1} \nonumber \\
&&\hspace{4cm}-\sum_{j=s+1}^n \frac{\aab{1}{s} \Delta^2(0,j;s)\Delta(0,j;s+1)}{D(0,j,q_{1,j})} \left(\sum_{k=1}^jz_k\right)\bigg],\\
\lefteqn{
 \ssplit(1_q^+,\ldots, s^{+}_{\bar Q},(s+1)^-_Q,\ldots,n^+ \to P_q^+)=
  \frac{1}{\sqrt{z_1 z_n} \prod_{l=1}^{n-1} \spa{l,}.{l+1} }}\nonumber \\
  &&\hspace{4cm}\times\bigg[
 \sum_{i=1}^{s-1} \sum_{j=s+1}^n
\frac{\Delta^3(i,j;s+1)\Delta(i,j;s)}{D(i,j,q_{i+1,j})} \sqrt{z_1} \nonumber \\
&&\hspace{4cm}-\sum_{j=s+1}^n \frac{\aab{1}{s} \Delta^3(0,j;s+1)}{D(0,j,q_{1,j})} \left(\sum_{k=1}^jz_k\right)\bigg].
\end{eqnarray}

Similarly, the two diagrams of $\Split_-(m_1,m_2)$-type yield,
\begin{eqnarray}
\lefteqn{
 \ssplit(1_q^-,\ldots, s^{-}_{\bar Q},(s+1)^+_Q,\ldots,n^+ \to P_q^-)=
  \frac{1}{\sqrt{z_1 z_n} \prod_{l=1}^{n-1} \spa{l,}.{l+1} }}\nonumber \\
  &&\hspace{4cm}\times\bigg[
  \sum_{i=1}^{s-1} \sum_{j=s+1}^n
\frac{\Delta(i,j;s+1)\Delta^3(i,j;s)}{D(i,j,q_{i+1,j})} \sqrt{z_1}^3 \nonumber \\
&&\hspace{4cm}-\sum_{j=s+1}^n \frac{\aab{1}{s}^3 \Delta(0,j;s+1)}{D(0,j,q_{1,j})}
  \left( \sum_{k=1}^jz_k \right)^3\bigg],\\
\lefteqn{
 \ssplit(1_q^-,\ldots, s^{+}_{\bar Q},(s+1)^-_Q,\ldots,n^+ \to P_q^-)=
  \frac{1}{\sqrt{z_1 z_n} \prod_{l=1}^{n-1} \spa{l,}.{l+1} }}\nonumber \\
  &&\hspace{4cm}\times\bigg[
  \sum_{i=1}^{s-1} \sum_{j=s+1}^n
\frac{\Delta^3(i,j;s+1)\Delta(i,j;s)}{D(i,j,q_{i+1,j})} \sqrt{z_1}^3 \nonumber\\
&&\hspace{4cm}-\sum_{j=s+1}^n \frac{\aab{1}{s}\aab{1}{s+1}^2 \Delta(0,j;s+1)}{D(0,j,q_{1,j})}
  \left( \sum_{k=1}^jz_k \right)^3 \bigg].
\end{eqnarray}

\subsection{Three quarks in the collinear set: $q (ng) \bar q Q \to Q$}

Here the relevant vertices in the MHV rules include
the $\tilde A$
four-quark amplitudes of eqs.~(\ref{eq:4qmhv5})--(\ref{eq:4qmhv8})
and the factorised amplitude is a two-quark MHV as given in  eq.~(\ref{eq:2qMHV})
As in the previous case, the quark helicities are constrained such that
there are no $\Delta M=0$ splitting functions.

\subsubsection{$\Delta M = 1$}

There are two diagrams for both $\Split_+(m_1)$- and $\Split_-(m_1,m_2)$-types
and we find,
\begin{eqnarray}
\lefteqn{
 \widetilde{\ssplit}(1_q^+,\ldots, s^{-}_{\bar q},(s+1)^+_Q,\ldots,n^+ \to P_Q^+)=
  \frac{1}{\sqrt{z_1 z_n} \prod_{l=1}^{n-1} \spa{l,}.{l+1} }}\nonumber \\
  &&\hspace{4cm}\times\bigg[
\frac{\Delta(0,s;1)\Delta^3(0,s;s)}{D(0,s,q_{1,s})} \sqrt{z_{s+1}} \nonumber \\
&&\hspace{4cm}+ \sum_{j=s+1}^n \frac{\aab{s}{s+1} \Delta^2(0,j;s)\Delta(0,j;1)}{D(0,j,q_{1,j})} \left(\sum_{k=1}^jz_k\right)\bigg],\\
\lefteqn{
 \widetilde{\ssplit}(1_q^-,\ldots, s^{+}_{\bar q},(s+1)^+_Q,\ldots,n^+ \to P_Q^+)=
  \frac{1}{\sqrt{z_1 z_n} \prod_{l=1}^{n-1} \spa{l,}.{l+1} }}\nonumber \\
  &&\hspace{4cm}\times\bigg[
\frac{\Delta^3(0,s;1)\Delta(0,s;s)}{D(0,s,q_{1,s})} \sqrt{z_{s+1}} \nonumber \\
&&\hspace{4cm}+ \sum_{j=s+1}^n \frac{\aab{s}{s+1} \Delta^3(0,j;1)}{D(0,j,q_{1,j})} \left(\sum_{k=1}^jz_k\right)\bigg],
\end{eqnarray}
and,
\begin{eqnarray}
\lefteqn{
 \widetilde{\ssplit}(1_q^+,\ldots, s^{-}_{\bar q},(s+1)^-_Q,\ldots,n^+ \to P_Q^-)=
  \frac{1}{\sqrt{z_1 z_n} \prod_{l=1}^{n-1} \spa{l,}.{l+1} }}\nonumber \\
  &&\hspace{4cm}\times\bigg[
\frac{\Delta^3(0,s;1)\Delta(0,s;s)}{D(0,s,q_{1,s})} \sqrt{z_{s+1}}^3 \nonumber \\
&&\hspace{4cm}+ \sum_{j=s+1}^n \frac{\aab{s}{s+1}^3 \Delta(0,j;1)}{D(0,j,q_{1,j})} \left(\sum_{k=1}^jz_k\right)^3\bigg],\\
\lefteqn{
 \widetilde{\ssplit}(1_q^-,\ldots, s^{+}_{\bar q},(s+1)^-_Q,\ldots,n^+ \to P_Q^-)=
  \frac{1}{\sqrt{z_1 z_n} \prod_{l=1}^{n-1} \spa{l,}.{l+1} }}\nonumber \\
  &&\hspace{4cm}\times\bigg[
\frac{\Delta^3(0,s;1)\Delta(0,s;s)}{D(0,s,q_{1,s})} \sqrt{z_{s+1}}^3 \nonumber \\
&&\hspace{4cm}+ \sum_{j=s+1}^n \frac{\aab{s}{s+1} \aab{1}{s+1}^2\Delta(0,j;1)}{D(0,j,q_{1,j})} \left(\sum_{k=1}^jz_k\right)^3
\bigg].
\end{eqnarray}

\subsection{Four quarks in the collinear set: $\bar Q Q (ng) \bar q q \to g$}

This limit is associated with the  four-quark $A$-type colour ordered amplitude and
is obtained by factoring onto a gluonic MHV.

\subsubsection{$\Delta M = 1$}

\begin{figure}[t!]
    \centering
    \psfrag{s+1}{$s+1$}
    \psfrag{s}{$s$}
    \psfrag{t+1}{$t+1$}
    \psfrag{t}{$t$}
    \begin{center}
        \includegraphics[width=12cm]{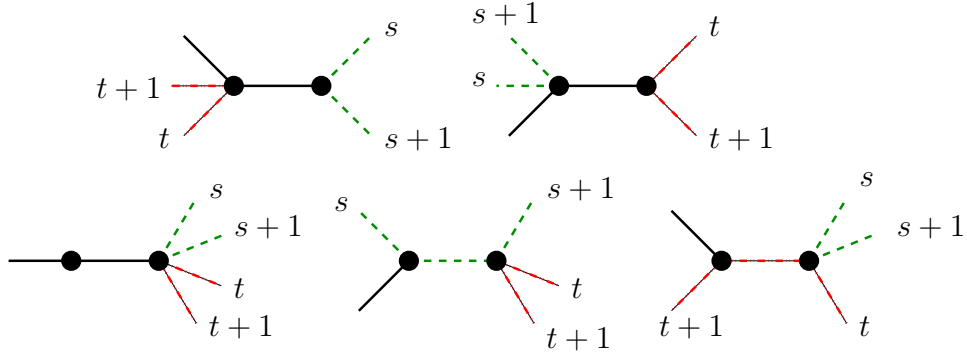}
    \end{center}
    \caption{MHV topologies contributing to the four quark collinear limit of the type
    $\ssplit(1^+,\ldots,s_{\bar Q}^\lambda, (s+1)_Q^{-\lambda}, \ldots t_{\bar q}^{\lambda^\prime},
  (t+1)_q^{-\lambda^\prime}, \ldots, n^+ \to P^-)$.
    Quarks of type $Q$ ($q$) are shown as green(red)-dotdashed lines and negative
    helicity gluons as black solid lines.
 }
 \label{fig:4q}
\end{figure}

Because of helicity
conservation for the quarks, $\Delta M = 0$ is forbidden. Furthermore, at least two
negative helicity quarks participate in the scattering so that $\Delta M = 1$
splittings must be of the $\Split_-(m_1,m_2)$-type.
The five contributing diagrams
are shown in Fig.~\ref{fig:4q}.
Explicit evaluation of the four independent helicity configurations yields,
\begin{eqnarray}
\lefteqn{
  \ssplit(1^+,\ldots,s_{\bar Q}^+, (s+1)_Q^-, \ldots t_{\bar q}^-,
  (t+1)_q^+, \ldots, n^+ \to P^-)=
  \frac{1}{\sqrt{z_1 z_n} \prod_{l=1}^{n-1} \spa{l,}.{l+1} }}\nonumber \\
 &&\hspace{4cm}\times\bigg[
\sum_{i=0}^{s-1}\sum_{j=s+1}^{t-1}
  \frac{\Delta(i,j;s)\Delta^3(i,j;s+1)}{D(i,j,q_{i+1,j})}
  \sqrt{z_t}^3\sqrt{z_{t+1}} \nonumber\\
&&\hspace{4cm}+ \sum_{i=s+1}^{t-1}\sum_{j=t+1}^n
  \frac{\Delta^3(i,j;t)\Delta(i,j;t+1)}{D(i,j,q_{i+1,j})}
  \sqrt{z_s}\sqrt{z_{s+1}}^3 \nonumber\\
&&\hspace{4cm}+ \sum_{i=0}^{s-1} \sum_{j=t+1}^n
  \frac{\aab{s+1}{t}^3\aab{s}{t+1}}{D(i,j,q_{i+1,j})}\left(\sum_{k=i+1}^j z_k\right)^4 \nonumber\\
&&\hspace{4cm}- \sum_{j=t+1}^n
  \frac{\Delta(s,j;t+1)\aab{s+1}{t}^3}{D(s,j,q_{s+1,j})}\sqrt{z_{s}}
  \left(\sum_{k=s+1}^j z_k\right)^3 \nonumber\\
&&\hspace{4cm}+  \sum_{i=0}^{s-1}
  \frac{\Delta(i,t;s)\aab{s+1}{t}^3}{D(i,t,q_{i+1,t})}\sqrt{z_{t+1}}
  \left( \sum_{k=i+1}^{t} z_k \right)^3 \bigg],
\\
\lefteqn{
  \ssplit(1^+,\ldots,s_{\bar Q}^-, (s+1)_Q^+, \ldots t_{\bar q}^-,
  (t+1)_q^+, \ldots, n^+ \to P^-)=
  \frac{1}{\sqrt{z_1 z_n} \prod_{l=1}^{n-1} \spa{l,}.{l+1} }}\nonumber \\
 &&\hspace{4cm}\times\bigg[
\sum_{i=0}^{s-1}\sum_{j=s+1}^{t-1}
  \frac{\Delta^3(i,j;s)\Delta(i,j;s+1)}{D(i,j,q_{i+1,j})}
  \sqrt{z_t}^3\sqrt{z_{t+1}} \nonumber\\
&&\hspace{4cm}+ \sum_{i=s+1}^{t-1}\sum_{j=t+1}^n
  \frac{\Delta^3(i,j;t)\Delta(i,j;t+1)}{D(i,j,q_{i+1,j})}
  \sqrt{z_s}^3\sqrt{z_{s+1}} \nonumber\\
&&\hspace{4cm}+ \sum_{i=0}^{s-1} \sum_{j=t+1}^n
  \frac{\aab{s+1}{t}\aab{s}{t+1}\aab{t}{s}^2}{D(i,j,q_{i+1,j})}\left(\sum_{k=i+1}^j z_k\right)^4 \nonumber\\
&&\hspace{4cm}- \sum_{j=t+1}^n
  \frac{\Delta(s,j;t+1)\Delta^2(s,j;t)\aab{s+1}{t}}{D(s,j,q_{s+1,j})}\sqrt{z_{s}}^3
  \left(\sum_{k=s+1}^j z_k\right) \nonumber\\
&&\hspace{4cm}+  \sum_{i=0}^{s-1}
  \frac{\Delta(i,t;s)\aab{s+1}{t}\aab{t}{s}^2}{D(i,t,q_{i+1,t})}\sqrt{z_{t+1}}
  \left( \sum_{k=i+1}^{t} z_k \right)^3 \bigg],
\\
\lefteqn{
  \ssplit(1^+,\ldots,s_{\bar Q}^+, (s+1)_Q^-, \ldots t_{\bar q}^+,
  (t+1)_q^-, \ldots, n^+ \to P^-)=
  \frac{1}{\sqrt{z_1 z_n} \prod_{l=1}^{n-1} \spa{l,}.{l+1} }}\nonumber \\
 &&\hspace{4cm}\times\bigg[
\sum_{i=0}^{s-1}\sum_{j=s+1}^{t-1}
  \frac{\Delta(i,j;s)\Delta^3(i,j;s+1)}{D(i,j,q_{i+1,j})}
  \sqrt{z_t}\sqrt{z_{t+1}}^3 \nonumber\\
&&\hspace{4cm}+ \sum_{i=s+1}^{t-1}\sum_{j=t+1}^n
  \frac{\Delta(i,j;t)\Delta^3(i,j;t+1)}{D(i,j,q_{i+1,j})}
  \sqrt{z_s}\sqrt{z_{s+1}}^3 \nonumber\\
&&\hspace{4cm}+ \sum_{i=0}^{s-1} \sum_{j=t+1}^n
  \frac{\aab{s+1}{t}\aab{s}{t+1}\aab{s+1}{t+1}^2}{D(i,j,q_{i+1,j})}\left(\sum_{k=i+1}^j z_k\right)^4 \nonumber\\
&&\hspace{4cm}- \sum_{j=t+1}^n
  \frac{\Delta(s,j;t+1)\aab{s+1}{t}\aab{s+1}{t+1}^2}{D(s,j,q_{s+1,j})}\sqrt{z_{s}}
  \left(\sum_{k=s+1}^j z_k\right)^3 \nonumber\\
&&\hspace{4cm}+  \sum_{i=0}^{s-1}
  \frac{\Delta(i,t;s)\Delta^2(i,t;s+1)\aab{s+1}{t}}{D(i,t,q_{i+1,t})}\sqrt{z_{t+1}}^3
  \left( \sum_{k=i+1}^{t} z_k \right)\bigg],\nonumber \\
\\
\lefteqn{
  \ssplit(1^+,\ldots,s_{\bar Q}^-, (s+1)_Q^+, \ldots t_{\bar q}^+,
  (t+1)_q^-, \ldots, n^+ \to P^-)=
  \frac{1}{\sqrt{z_1 z_n} \prod_{l=1}^{n-1} \spa{l,}.{l+1} }}\nonumber \\
 &&\hspace{4cm}\times\bigg[
\sum_{i=0}^{s-1}\sum_{j=s+1}^{t-1}
  \frac{\Delta^3(i,j;s)\Delta(i,j;s+1)}{D(i,j,q_{i+1,j})}
  \sqrt{z_t}\sqrt{z_{t+1}}^3 \nonumber\\
&&\hspace{4cm}+ \sum_{i=s+1}^{t-1}\sum_{j=t+1}^n
  \frac{\Delta(i,j;t)\Delta^3(i,j;t+1)}{D(i,j,q_{i+1,j})}
  \sqrt{z_s}^3\sqrt{z_{s+1}} \nonumber\\
&&\hspace{4cm}+ \sum_{i=0}^{s-1} \sum_{j=t+1}^n
  \frac{\aab{s+1}{t}\aab{s}{t+1}^3}{D(i,j,q_{i+1,j})}\left(\sum_{k=i+1}^j z_k\right)^4 \nonumber\\
&&\hspace{4cm}- \sum_{j=t+1}^n
  \frac{\Delta^3(s,j;t+1)\aab{s+1}{t}}{D(s,j,q_{s+1,j})}\sqrt{z_{s}}^3
  \left(\sum_{k=s+1}^j z_k\right) \nonumber\\
&&\hspace{4cm}+  \sum_{i=0}^{s-1}
  \frac{\Delta^3(i,t;s)\aab{s+1}{t}}{D(i,t,q_{i+1,t})}\sqrt{z_{t+1}}^3
  \left( \sum_{k=i+1}^{t} z_k \right) \bigg].
\end{eqnarray}

\subsection{Four quarks in the collinear set: $\bar Q q (ng) \bar q Q \to g$}

This limit is associated with the  four-quark $\tilde A$-type colour ordered
 amplitude and
is obtained by factoring onto a gluonic MHV.

\subsubsection{$\Delta M = 1$}

\begin{figure}[t!]
    \centering
    \psfrag{s+1}{$s+1$}
    \psfrag{s}{$s$}
    \psfrag{t+1}{$t+1$}
    \psfrag{t}{$t$}
    \begin{center}
        \includegraphics[height=5cm]{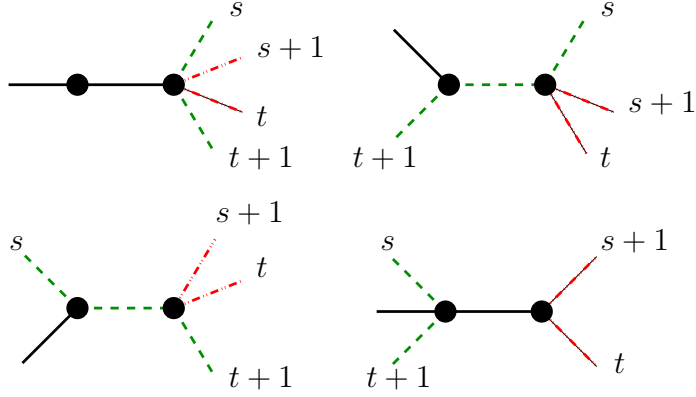}
    \end{center}
    \caption{MHV topologies contributing to the four quark collinear limit of the type
    $\widetilde{\ssplit}(1^+,\ldots,s_{\bar Q}^\lambda, (s+1)_q^{-\lambda^\prime}, \ldots t_{\bar q}^{\lambda^\prime},
  (t+1)_Q^{-\lambda}, \ldots, n^+ \to P^-)$.
    Quarks of type $Q$ ($q$) are shown as green(red)-dotdashed lines and negative
    helicity gluons as black solid lines.
 }
\label{fig:4qtilde}
\end{figure}

As in the previous case,  helicity
conservation for the quarks, ensures that $\Delta M = 0$ is forbidden and
that $\Split_+(m_1)$-type $\Delta M = 1$ splittings are absent.
The four  contributing diagrams of $\Split_-(m_1,m_2)$-type
are shown in Fig.~\ref{fig:4qtilde}.

The four independent helicity configurations are given by,
\begin{eqnarray}
\lefteqn{
  \widetilde{\ssplit}(1^+,\ldots,s_{\bar Q}^+, (s+1)_q^+, \ldots t_{\bar q}^-,
  (t+1)_Q^-, \ldots, n^+ \to P^-)=
  \frac{1}{\sqrt{z_1 z_n} \prod_{l=1}^{n-1} \spa{l,}.{l+1} }}\nonumber \\
 &&\hspace{5cm}\times\bigg[
\sum_{i=0}^{s-1} \sum_{j=t+1}^{n}
  \frac{\aab{t+1}{t}^3\aab{s}{s+1}}{D(i,j,q_{i+1,j})} \left( \sum_{k=i+1}^j  z_k\right)^4
  \nonumber\\
&&\hspace{5cm}- \sum_{j=t+1}^n \frac{\Delta(s,j;s+1)\aab{t+1}{t}^3}{D(s,j,q_{s+1,j})}
  \sqrt{z_s} \left( \sum_{k=s+1}^j z_k \right)^3 \nonumber\\
&&\hspace{5cm}- \sum_{i=0}^{s-1} \frac{\Delta^3(i,t;t)\aab{s}{s+1}}{D(i,t,q_{i+1,t})}
  \sqrt{z_{t+1}}^3 \left( \sum_{k=i+1}^t z_k \right) \nonumber\\
&&\hspace{5cm}+ \frac{\Delta^3(s,t;t)\Delta(s,t;s+1)}{D(s,t,q_{s+1,t})}\sqrt{z_s}\sqrt{z_{t+1}}^3\bigg],
\\
\lefteqn{
  \widetilde{\ssplit}(1^+,\ldots,s_{\bar Q}^+, (s+1)_q^-, \ldots t_{\bar q}^+,
  (t+1)_Q^-, \ldots, n^+ \to P^-)=
  \frac{1}{\sqrt{z_1 z_n} \prod_{l=1}^{n-1} \spa{l,}.{l+1} }}\nonumber \\
 &&\hspace{5cm}\times\bigg[
\sum_{i=0}^{s-1} \sum_{j=t+1}^{n}
  \frac{\aab{t+1}{t}\aab{s}{s+1}\aab{s+1}{t+1}^2}{D(i,j,q_{i+1,j})} \left( \sum_{k=i+1}^j  z_k\right)^4
  \nonumber\\
&&\hspace{5cm}- \sum_{j=t+1}^n \frac{\Delta(s,j;s+1)\aab{t+1}{t}\aab{s+1}{t+1}^2}{D(s,j,q_{s+1,j})}
  \sqrt{z_s} \left( \sum_{k=s+1}^j z_k \right)^3 \nonumber\\
&&\hspace{5cm}- \sum_{i=0}^{s-1} \frac{\Delta(i,t;t)\Delta^2(i,t;s+1)\aab{s}{s+1}}{D(i,t,q_{i+1,t})}
  \sqrt{z_{t+1}}^3 \left( \sum_{k=i+1}^t z_k \right) \nonumber\\
&&\hspace{5cm}+ \frac{\Delta(s,t;t)\Delta^3(s,t;s+1)}{D(s,t,q_{s+1,t})}\sqrt{z_s}\sqrt{z_{t+1}}^3\bigg],
\\
\lefteqn{
  \widetilde{\ssplit}(1^+,\ldots,s_{\bar Q}^-, (s+1)_q^+, \ldots t_{\bar q}^-,
  (t+1)_Q^+, \ldots, n^+ \to P^-)=
  \frac{1}{\sqrt{z_1 z_n} \prod_{l=1}^{n-1} \spa{l,}.{l+1} }}\nonumber \\
 &&\hspace{5cm}\times\bigg[
\sum_{i=0}^{s-1} \sum_{j=t+1}^{n}
  \frac{\aab{t+1}{t}\aab{s}{s+1}\aab{t}{s}^2}{D(i,j,q_{i+1,j})} \left( \sum_{k=i+1}^j  z_k\right)^4
  \nonumber\\
&&\hspace{5cm}- \sum_{j=t+1}^n \frac{\Delta(s,j;s+1)\Delta^2(s,j;t)\aab{t+1}{t}}{D(s,j,q_{s+1,j})}
  \sqrt{z_s}^3 \left( \sum_{k=s+1}^j z_k \right) \nonumber\\
&&\hspace{5cm}- \sum_{i=0}^{s-1} \frac{\Delta(i,t;t)\aab{s}{s+1}\aab{t}{s}^2}{D(i,t,q_{i+1,t})}
  \sqrt{z_{t+1}} \left( \sum_{k=i+1}^t z_k \right)^3 \nonumber \\
&&\hspace{5cm}+ \frac{\Delta^3(s,t;t)\Delta(s,t;s+1)}{D(s,t,q_{s+1,t})}\sqrt{z_s}^3\sqrt{z_{t+1}}\bigg],
\\
\lefteqn{
  \widetilde{\ssplit}(1^+,\ldots,s_{\bar Q}^-, (s+1)_q^-, \ldots t_{\bar q}^+,
  (t+1)_Q^+, \ldots, n^+ \to P^-)=
  \frac{1}{\sqrt{z_1 z_n} \prod_{l=1}^{n-1} \spa{l,}.{l+1} }}\nonumber \\
 &&\hspace{5cm}\times\bigg[
\sum_{i=0}^{s-1} \sum_{j=t+1}^{n}
  \frac{\aab{t+1}{t}\aab{s}{s+1}^3}{D(i,j,q_{i+1,j})} \left( \sum_{k=i+1}^j  z_k\right)^4
  \nonumber\\
&&\hspace{5cm}- \sum_{j=t+1}^n \frac{\Delta^3(s,j;s+1)\aab{t+1}{t}}{D(s,j,q_{s+1,j})}
  \sqrt{z_s}^3 \left( \sum_{k=s+1}^j z_k \right) \nonumber\\
&&\hspace{5cm}- \sum_{i=0}^{s-1} \frac{\Delta(i,t;t)\aab{s}{s+1}^3}{D(i,t,q_{i+1,t})}
  \sqrt{z_{t+1}} \left( \sum_{k=i+1}^t z_k \right)^3 \nonumber\\
&&\hspace{5cm}+ \frac{\Delta(s,t;t)\Delta^3(s,t;s+1)}{D(s,t,q_{s+1,t})}\sqrt{z_s}^3\sqrt{z_{t+1}}\bigg].
\end{eqnarray}

\section{Selected results for triple collinear limits}
\label{sec:specific-results}

To illustrate our general results for multi-collinear limits,
in this section we list some of the
triple-collinear splitting functions. The
$\Delta M = 0$  splitting amplitudes are obtained directly from
MHV amplitudes and we do not list them here.   Explicit results are given in
Section~\ref{sec:general-results}.
 For the $\Delta M = 1$ (and therefore $\Delta P = 1$)
 amplitudes, there are two types of splitting function corresponding to
$\Split_{+}(m_1)$ and $\Split_{-}(m_1,m_2)$.   In the specific case of
three collinear particles, these are related by parity thereby reducing the
number of independent amplitudes to at most
two for each splitting.   Here, we
list only the most compact form of the amplitudes.

\subsection{$q g g \to q$}
There are only two independent $\Delta M = 1$ splitting amplitudes,
which can be obtained by setting $m=n=3$ in
Eqs.~(\ref{eq:qgg1}) and (\ref{eq:qgg2}).  Explicitly we find,
\begin{eqnarray}
\ssplit(1_q^+,2^+,3^- \to P_q^+)&=&
-{\frac {\spa{2}.{3}{z_{{2}}}^{3/2}}
{\sqrt {z_{{3}}}  s_{{2,3}} 
\left( \spa{2}.{1}\sqrt{z_{{2}}}+\spa{3}.{1}\sqrt {z_{{3}}} \right) 
\left( z_{{2}}+z_{{3}} \right) }}\nonumber \\
&&-{\frac {\spa{1}.{3} 
\left( \spa{1}.{3}\sqrt {z_{{1}}}+\spa{2}.{3}\sqrt {z_{{2}}} \right) ^{2}}
{\spa{1}.{2}\spa{2}.{3}s_{{1,3}} 
\left( \spa{2}.{1}\sqrt {z_{{2}}}+\spa{3}.{1}\sqrt {z_{{3}}} \right) }} ,\\
\ssplit(1_q^-,2^+,3^-\to P_q-)&=&
-{\frac {z_{{1}}\spa{2}.{3}{z_{{2}}}^{3/2}}
{\sqrt {z_{{3}}}s_{{2,3}}
 \left( \spa{2}.{1}\sqrt {z_{{2}}}+\spa{3}.{1}\sqrt {z_{{3}}} \right)
 \left( z_{{2}}+z_{{3}} \right) }}\nonumber \\
&&-{\frac {  \spa{1}.{3}^{3}}
{\spa{1}.{2}\spa{2}.{3}s_{{1,3}} 
\left( \spa{2}.{1}\sqrt {z_{{2}}}+\spa{3}.{1}\sqrt {z_{{3}}} \right) 
}}.
\end{eqnarray}
All others can be obtained by parity and charge conjugation.
These expressions numerically agree with the splitting functions
given in \cite{delduca}\footnote{Note that there is a small 
typographical error 
in Eq. (5.56) of Ref.~\cite{delduca}.  $s_{23}$ should be replaced by $s_{12}$
in the last term  of the equation for ${\rm split}_-^{q\to
qgg}(k_1^+,k_2^+,k_3^-)$},
\begin{eqnarray}
\ssplit(1_q^+,2^+,3^- \to P_q^+)&=&-\frac{1}{s_{1,2}s_{2,3}}\nonumber \\
&\times&
\bigg[
\frac{\spb{1}.{2}(\spa{1}.{3}\sqrt{z_1}+\spa{2}.{3}\sqrt{z_2})^2
(\spb{1}.{2}\sqrt{z_1}+\spb{3}.{2}\sqrt{z_3})}{s_{1,3}}\nonumber \\
&&
+\frac{\sqrt{z_2}(z_1+z_2)
\spb{1}.{2}(\spa{1}.{3}\sqrt{z_1}+\spa{2}.{3}\sqrt{z_2})}{\sqrt{z_3}}\nonumber \\
&&+\frac{\sqrt{z_1}z_2s_{1,2}}{(z_2+z_3)}\bigg], \\
\ssplit(1_q^-,2^+,3^-\to P_q-)&=&-\frac{1}{s_{1,2}s_{2,3}}\nonumber \\
&\times&
\bigg[
\frac{\spb{1}.{3}(\spa{1}.{2}\sqrt{z_1}+\spa{3}.{2}\sqrt{z_3})^2
(\spb{1}.{3}\sqrt{z_1}+\spb{2}.{3}\sqrt{z_2})}{s_{1,3}}\nonumber \\
&&
+\frac{\sqrt{z_2}(z_1+z_2)
\spb{1}.{3}(\spa{1}.{2}\sqrt{z_1}+\spa{3}.{2}\sqrt{z_3})}{\sqrt{z_3}}\nonumber \\
&&+\frac{\sqrt{z_1}z_2s_{1,2}}{(z_2+z_3)}+\sqrt{z_2}\spa{2}.{3}\spb{1}.{3}\bigg].
\end{eqnarray}
We see that the two sets of results have the same types of
singularity structure as $z_3 \to 0$ and $z_1 \to 1$ corresponding to the soft
and double soft gluon limits.

\subsection{$\bar q q g \to g$}
There are again only two independent $\Delta M = 1$ amplitudes,
both of which can be obtained from Eq.~(\ref{eq:gqq})
by setting $s=2$ and $\lambda=\pm\frac{1}{2}$.
  We find,
\begin{eqnarray}
\label{eq:qggppm}  
\ssplit(1^+,2_{\bar q}^+,3_q^-\to P^+)&=&-
{\frac { \left( \spa{1}.{2}\sqrt {z_{{1}}}+\spa{3}.{2}\sqrt {z_{{3}}}
 \right)  \left( \spa{1}.{3}\sqrt {z_{{1}}}+\spa{2}.{3}\sqrt {z_{{2}}}
 \right) ^{2}}{\spa{1}.{2}\spa{2}.{3}s_{{1,3}} \left( \spa{2}.{1}
\sqrt {z_{{2}}}+\spa{3}.{1}\sqrt {z_{{3}}} \right) }}\nonumber\\
&&+{\frac {z_{{2}}\spa{2}.
{3}}{\sqrt {z_{{1}}}s_{{2,3}} \left( \spa{2}.{1}\sqrt {z_{{2}}}
+\spa{3}.{1}\sqrt {z_{{3}}} \right)  \left( z_{{2}}+z_{{3}} \right) }}
,\\
\label{eq:qggpmp}  
\ssplit(1^+,2_{\bar q}^-,3_{q}^+\to P^+)&=&
-{\frac { 
\left( \spa{1}.{2}\sqrt {z_{{1}}}+\spa{3}.{2}\sqrt {z_{{3}}} \right)^{3}}
{\spa{1}.{2}\spa{2}.{3}s_{{1,3}}
\left( \spa{2}.{1}\sqrt {z_{{2}}}+\spa{3}.{1}\sqrt {z_{{3}}} \right)
}}\nonumber \\&&
+{\frac {z_{{3}}\spa{2}.{3}}{\sqrt {z_{{1}}}s_{{2,3}} 
\left( \spa{2}.{1}\sqrt {z_{{2}}}+\spa{3}.{1}\sqrt {z_{{3}}} \right)
\left( z_{{2}}+z_{{3}} \right) }}.
\end{eqnarray}
All others can be obtained via parity and charge
conjugation.
Eq.~(\ref{eq:qggppm}) numerically agrees with 
the analogous
expression given in Ref.~\cite{delduca}.
We were not able to find agreement  between  Eq.~(\ref{eq:qggpmp})
and the expression for
${\rm split}_-^{g \to g \bar q q}(k_1^+,k_2^-,k_3^+)$ in Eq.~(5.53)
of Ref.~\cite{delduca}.

\subsection{$q\bar Q Q \to q$ and $q\bar q Q \to Q$}
In this special case both colour structures lead to the same splitting
amplitude. There  is only one 
independent $\Delta M = 1$ helicity configurations.
It is given in Eq.~(\ref{eq:qpQpQm1}).

\section{Conclusion}
\label{sec:conclusion}

In this paper we have considered the collinear limit of multi-parton QCD
amplitudes at tree level. We have used the new MHV rules for constructing
colour ordered amplitudes from MHV vertices  to generalise our previous results
for gluon-only splitting functions. Our main results are general formulae for
timelike splitting functions involving up to two negative helicity partons and
an arbitrary number of positive helicity partons. We anticipate that the
expressions presented here will be useful in 
developing higher order perturbative
predictions for observable quantities, such as jet cross sections at the LHC
or in examining the high energy limit of QCD. 

A key point of our approach is that in the collinear limit only a
subset of MHV rules diagrams contribute - those where every propagator invariant
$s$  goes on-shell in the multi-collinear limit.
We observe that the splitting functions have a simple structure, 
and can be written as
sums over the corresponding poles in $s$ multiplied
by a coefficient that is either entirely composed of
holomorphic spinor products $\spa{i}.{j}$ or entirely composed of
anti-holomorphic spinor products $\spb{i}.{j}$.
This implies that the coefficients are
supported on a single degree-one curve in (anti)-twistor space.

\bigskip

\section{Acknowledgements}
EWNG and VVK acknowledge the support of PPARC through
Senior Fellowships and TGB acknowledges the award of a PPARC studentship.

\bibliographystyle{JHEP-2}
\bibliography{paper.bbl}

\end{document}